\documentclass[reprint,
amsmath,amssymb,
aps,
pra,
]{revtex4-1}
\usepackage{graphicx}
\usepackage{dcolumn}
\usepackage{bm}
\usepackage{hyperref}
\usepackage{url}
\usepackage{float}
\usepackage{subfigure}
\usepackage{setspace}
\begin{document}
\title{Time scales in stock markets}
\author{Ajit Mahata$^1$, Md Nurujjaman$^2$}
\email{$^2$jaman\_nonlinear@yahoo.co.in}
\affiliation{Department of Physics, NIT Sikkim, Ravangla, South Sikkim-737139, India}

\date{\today}

\begin{abstract}
Different investment strategies are adopted in short-term  and long-term depending on the time scales, even though time scales are adhoc in nature. Empirical mode decomposition based Hurst exponent analysis and variance technique have been applied to identify the time scales for short-term  and long-term investment from the decomposed intrinsic mode functions(IMF). Hurst exponent ($H$) is around 0.5 for the IMFs with time scales from few days to 3 months, and $H\geq0.75$ for the IMFs with the time scales $\geq5$ months. Short term time series [$X_{ST}(t)$] with time scales from few days to 3 months and $H~0.5$  and long term time series [$X_{LT}(t)$] with time scales $\geq5$ and $H\geq0.75$, which represent the dynamics of the market, are constructed from the IMFs.  The $X_{ST}(t)$ and $X_{LT}(t)$ show that the market is random in short-term and correlated in long term. The study also show that the $X_{LT}(t)$ is correlated with fundamentals of the company. The analysis will be useful for investors to design the investment and trading strategy.

\end{abstract}

\maketitle

Stock market is a complex dynamical system where evolution of the dynamics depend on the participation of different types of investor or traders~\cite{mantegna1999introduction,bouchaud2003theory,huang2003applications}. Investors in a stock market participate to make gain, and they implement different investment strategies depending on different investment time horizon (ITH)~\cite{peters1994fractal,jegadeesh1993returns,PhysRevE18-jean-Ph}. Traders can simultaneously trade in a particular stock frequently for short-term gain or infrequently for long-term investment. Participation of diversified investors in terms of ITH, reaction to information and purpose of investment are very much important to get stabilized markets~\cite{peters1994fractal}.

It has been observed from market participation of the traders that the ITH of a short-term trader ranges from single day to few months, whereas a long-term trader invests with an ITH from few months to several years~\cite{kristoufek2012fractal}. Survey results on the investment techniques used by the several fund managers and foreign exchange dealers of various countries show that the technical analysis is used for short-term investment of ITH of day to few months, and the fundamental analysis is used for the long-term investment of ITH of more than few months to several years~\cite{menkhoff2010use,lui1998use}. These survey also show that the short-term and long-term market dynamics are mostly controlled by the psychological behaviour of the investors and  the fundamentals of the markets respectively~\cite{menkhoff2010use}. In these works the time scales for short-term and long-term ITH are defined on the investment experience and ad-hoc in nature, even though the separation of the short-term and long-term dynamics in terms of time scales is very important for the prediction of future price movement. However, no such study has been carried out so far to identify the time scales for short-term ITH and long-term ITH.

In this letter, we identify the existence of time scales that characterizes the dynamics of the market in short-term and long-term investment horizon by analyzing twelve leading global stock indices and stock price of some companies. The two distinct time horizons have been obtained based on the nature of correlations that has been quantified by estimating the Hurst exponent of the decomposed time series. Finally, two distinct time series of two different time horizons have been constructed from the stock indices and price. The time series with the time horizon with few days to 3 months is random in nature, and the other time series with time horizon greater than 5 months shows long range correlation. The second reconstructed time series, which is found to be correlated with the fundamentals of the companies, can be used to predict the future price movement.

The stock indices have been decomposed by using empirical mode decomposition (EMD) method, which preserves the nonstationarity and nonlinearity of a signal, in various monofrequency intrinsic mode function (IMF) of different time scales~\cite{huang1998empirical,huang1998engineering}. The IMFs satisfy the following two conditions ($i$) the number of extrema and the number of zero crossing must be equal or differ by one; and ($ii$) mean values of the envelope defined by the local maxima and local minima for each point is zero. The IMF is calculated in the following way:
(a) lower envelope $U(t)$ and upper envelope $V(t)$ are drawn by connecting minima and maxima of the data respectively using spline fitting. 
(b) Mean value of the envelope $m=[U(t)+V(t)]/2$ is subtracted from the original time series to get new data set $h= X(t)-m.$
(c) Repeat the process (a) and (b) by considering $h$ as a new data set until the IMF conditions ($i$ \& $ii$) are satisfied. Once the conditions  are satisfied, the process terminates and $h$ is stored as first IMF. The second IMF is calculated repeating the above steps (a)-(c) from the data set $d(t)=X(t)-IMF1$. When the final residual is monotonic in nature, the steps (a)-(c) are terminated and the orignal time series can be written as a set of IMFs plus trend, $X(t)=\displaystyle\sum_{i=1}^n IMF_i+residual,$ where $IMF_i$ represents the $i^{th}$ IMF. 

Each IMF represents a signal with particular time scale. The IMF1 contains the lowest time scale present in the time series, and the IMF2 contains the second lowest time scale and so on. It can be concluded that the IMF1 fluctuates faster than the IMF2 and so on. Hence, EMD technique can be used to separate various important time scales present in a signal in the form of IMFs. The characteristic time scale ($\tau$) of each IMF can be estimated from the frequency ($\omega$) by using Hilbert Transform (HT), which is define as $\displaystyle Y(t)=\frac{P}{\pi}\int_{-\infty}^{\infty}\frac{IMF(t)}{t-t'}dt$, where $P$ is the Cauchy principle value, and $\tau=1/\omega$ where $\omega=\frac{d\theta(t)}{dt}$, and $\displaystyle \theta(t)=tan^{-1}\frac{Y(t)}{IMF(t)}$~\cite{huang1998empirical}. Identification of important IMFs are very essential to separate the market dynamics in terms of short-term and long-term ITH, and it can be done by evaluating the Hurst exponent ($H$) of the IMFs. 

The Hurst exponent is estimated using rescaled range analysis (R/S) technique~\cite{hurst1951long}. For that we construct a time series $X_p$ defined as $Z_i=\displaystyle\sum_{p=1}^i(X_p-\bar{X})$, where $\bar{X}$ represents the average value of $X_p$. Now estimate the ratio of the rescaled range ($R$) over the standard deviation ($S$) of the $Z_i$ in various scales $l$. The ratio of each partial time series of length $l$ can be expressed as $(R/S)\propto l^H$, where $H$ is the Hurst exponent. For a random time series, $H$ is around $0.5$, and for correlated and anti-correlated time series, $H$ is greater than $0.5$, and less than $0.5$ respectively.
   
We have analysed the stock market indices from December 1995 to July 2018 of (1) S\&P 500 (USA), (2) Nikkei 225 (Japan), (3) CAC 40 (France), (4) IBEX 35 (Spain) (5) HSI (Hong Kong), (6) SSE (China), (7) BSE SENSEX (India), (8) IBOVESPA (Brazil), (9) BEL 20 (Euro-Next Brussels), (10) IPC (Mexico), (11) Russel2000(London),(12) TA125 (Israel) and stock price of the company (13) IBM (USA), (14) Microsoft (USA), (15) Tata Motors (India), (16) Reliance Communication (RCOM) (India), (17) Apple (USA) and (18) Reliance Industries Limited (RIL) (India) ~\cite{yahoo}.
 
\begin{figure}
\includegraphics[angle=0, width=8cm]{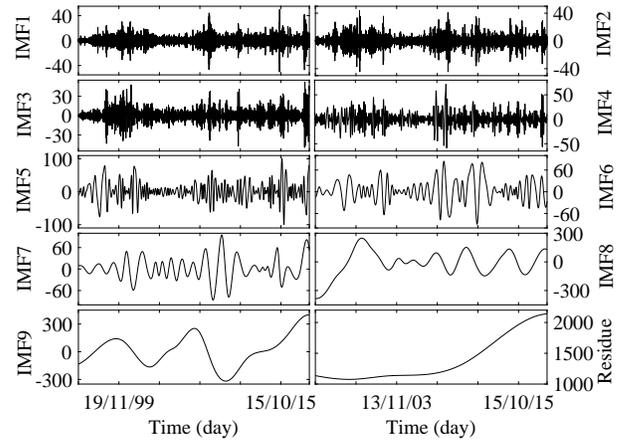}
\caption{\label{fig:SP_IMF}The plots one to nine and last one represent the IMFs and residue respectively of the S\&P 500 index.}
\end{figure}

Fig.~\ref{fig:SP_IMF} shows the IMF1 to IMF9 and the residue of the S\&P 500 index. IMF1 in Fig.~\ref{fig:SP_IMF} represents the mode with lowest time scale, and it gradually increases with the increase of IMF numbers. The residue represents the trend of the index.

\begin{figure}
\includegraphics[angle=0, width=8.4cm]{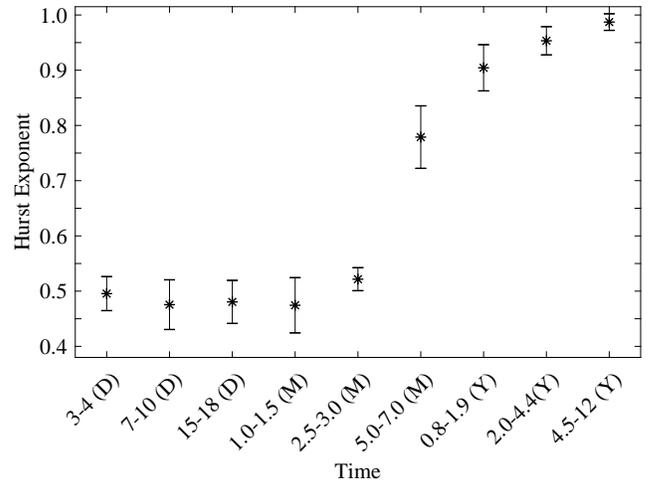}
\caption{\label{fig:Hurst} Hurst exponent ($H$) vs timescales of all the IMFs of all the indices and companies with $2\sigma$ error bar. The first point represents the average value of $H$ of all the first IMFs of all stock data, the second point represents the average value of $H$ of all the second IMFs of all stock data and so on. $H$ is around 0.5 upto IMF5 with a maximum time scale of around 3 months. The value of $H$ jumps to 0.75 for IMF6 (with a time scale 5 months) and gradually increases for IMF7 to IMF9. $H$ value shows that nature of IMF1 to IMF5 is random and IMF6 to IMF9 has a long range correlation. (D), (M) and (Y) in the $x$-axis represent the day, month and year respectively.}
\end{figure}

Fig.~\ref{fig:Hurst} shows $H$ of all the IMFs for all the market indices and stock price data. The value of $H~0.5$ for IMF1 to IMF5 with time scales from few days to 3 months. The value of $H$ jumps to 0.75 for IMF6 with a starting time scale of around 5 months. The value of $H$ gradually increases for IMF7 to IMF9 with a time scale ranging from 1 year to 12 years. The value of $H=0.5$ for IMF1 to IMF5 indicates that the nature of first five IMFs is random. Time scales of IMF1, IMF2, IMF3, IMF4 and IMF5 of all the stock data analysed here are in the range of 3-4 days, 7-10 days,  15-18 days, 1-1.5 months and 2.5-3 months respectively.

\begin{figure}
\includegraphics[angle=0, width=8.5cm]{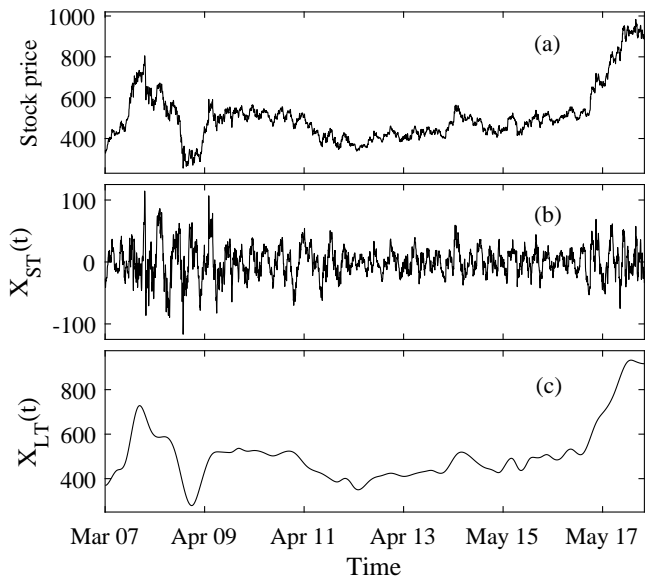}
\caption{\label{fig:RIL_funda}(a) Represents the original data of RIL from March 2007 to March 2018. Fig. (b) and (c) represent reconstructed $X_{ST}(t)$ and $X_{LT}(t)$ respectively.}
\end{figure}

We have constructed a time series ($X_{ST}(t)$), which is random in nature, by adding the IMF1 to IMF5, i.e. $X_{ST}(t)=\displaystyle\sum_{i=1}^5~IMF_i$. The Fig.~\ref{fig:RIL_funda}(b) shows the reconstructed time seires $X_{ST}(t)$ from original time series of RIL shown in Fig.~\ref{fig:RIL_funda}(a). The time series $X_{ST}(t)$ shows that the stock market is random in nature with the time scales ranging from few days to 3 months, and hence this range represents the time scales for the short-term ITH. From the analyses one can conclude that the ITH of few days to 3 months will be random in nature. As the technical analysis is usually applied to identify various trend pattern in this ITH, which is mainly depends on the investors' psychological behaviors~\cite{menkhoff2010use}, there may be some mean reversing short term trends in this time scales though overall dynamics is random in nature. Hence technical forecast cannot be simulated~\cite{allen1990charts}. Technical pattern that can be identified from the $X_{ST}(t)$ will be presented elsewhere.

The values of $H\geq 0.75$ for IMF6 to IMF9 indicate that there is long-range correlation in IMF6 to IMF9. The time scales of IMF6, IMF7, IMF8and IMF9 of all the stock data analysed here are in the range of 5-7 months, 0.8-1.9 yrs, 2-4.4 yrs, 4.5-12 yrs respectively. We can construct a time series ($X_{LT}(t)$) by adding the IMF6 to IMF9 and residue, i.e., $\displaystyle X_{LT}(t)=[\displaystyle\sum_{i=6}^9~IMF_i+Residue]$. The Fig.~\ref{fig:RIL_funda}(c) shows the reconstructed time series $X_{ST}(t)$ from original time series of RIL shown in Fig.~\ref{fig:RIL_funda}(a). The reconstructed time series $X_{LT}(t)$ represents the dynamics of the stock market with the time scales ranging from 5 months to few years and hence this range represents the time scales for the long-term ITH. From these analysis one can conclude that the ITH more than few months has long-range correlation, and hence  may be used to predict future price. One can reconstruct $X_{ST}(t)$ and $X_{LT}(t)$ for all the indies and stocks.  The study of correlation coefficients between $X_{LT}(t)$ and fundamentals of the companies is given below.

Table~\ref{tab:corr} shows that the correlation coefficient between $X_{LT}(t)$ and three fundamental variables: sale, net profit (NP) and cash from operating activity (COA) for 14 companies which are listed in NSE SENSEX and 6 companies which are listed in NASDAQ and NYSE from March 2007 to March 2018 in annual price level. We obtained positive correlation between  $X_{LT}(t)$ and sale, NP and COA for all years. It implies that stock price is highly correlated with sale, NP and COA. Hence one can conclude that for long-term ITH, fundamentals of a company are the most important parameters for prediction of future price.

\begin{table}
\caption{\label{tab:corr} Correlation coefficient between $X_{LT}(t)$ and three fundamental variables of some Indian and American companies. First column: sale, second Column: NP and third column: COA.}
\begin{ruledtabular}
\begin{tabular}{lrrr}
Company & Sale & NP & COA \\
Code &  & &   \\
\hline
NSE: ASIANPAINT & 0.9930 & 1.0000 & 0.9441 \\
NSE: BPCL & 0.6923 & 0.8322 & 0.6853 \\
NSE: COPLA & 0.8951 & 0.6713 & 0.6713 \\
NSE: DRREDDY & 0.9441 & 0.8671 & 0.9510 \\
NSE: EICHERMOT & 0.9441 & 0.9790 & 0.9860 \\
NSE: GAIL & 0.5664 & 0.5804 & 0.5804 \\
NSE: GRASIM & 0.8462 & 0.4615 & 0.4615 \\
NSE: HCLTECH & 0.9441 & 0.9231 & 0.8951 \\
NSE: HEROMOTOCO & 0.9650 & 0.9580 & 0.9021 \\
NSE: HINDALCO & 0.2587 & 0.4545 & 0.3776 \\
NSE: HINDUNILVR & 0.9720 & 0.9860 & 0.8252 \\
NSE: TATAMOTORS & 0.8462 & 0.7832 & 0.9091 \\
NSE: RCOM  & 0.0490 & 0.9510 & 0.5035 \\
NSE: RELIANCE & 0.1119 & 0.7203 & 0.4476 \\
NYSE: JNJ & 0.8352 & 0.2747 & 0.8022 \\
NASDAQ: AMZN & 0.9890 & 0.4231 & 0.9890 \\
NASDAQ: GOOGL & 0.9231 & 0.8462 & 0.9231 \\
NASDAQ: AAPL & 0.9615 & 0.9396 & 0.9341 \\
NASDAQ: MSFT & 0.7253 & 0.0659 & 0.6484 \\
NASDAQ: INTC & 0.9011 & 0.5879 & 0.8681 \\
\end{tabular}
\end{ruledtabular}
\end{table}

To further verify the robustness of the proposed method, analysis of the decomposed time series has been carried out using Normalised Variance ($NV$) technique. Based on the NV technique, we can identify the important IMF. The technique estimates the energy of the $i^{th}$ IMFs by calculating variance~~\cite{chatlani2012emd,zao2014speech}, and the $NV$ of $i^{th}$ is defined as $\displaystyle NV_i=\frac{\sqrt{\displaystyle\sum_t IMF_i^2(t)}}{\displaystyle\sum_{i=1}^N\sqrt{\displaystyle\sum_t IMF_i^2(t)}},$ where, $N$ is the total number of IMF.

\begin{figure}
\includegraphics[angle=0, width=8.4cm]{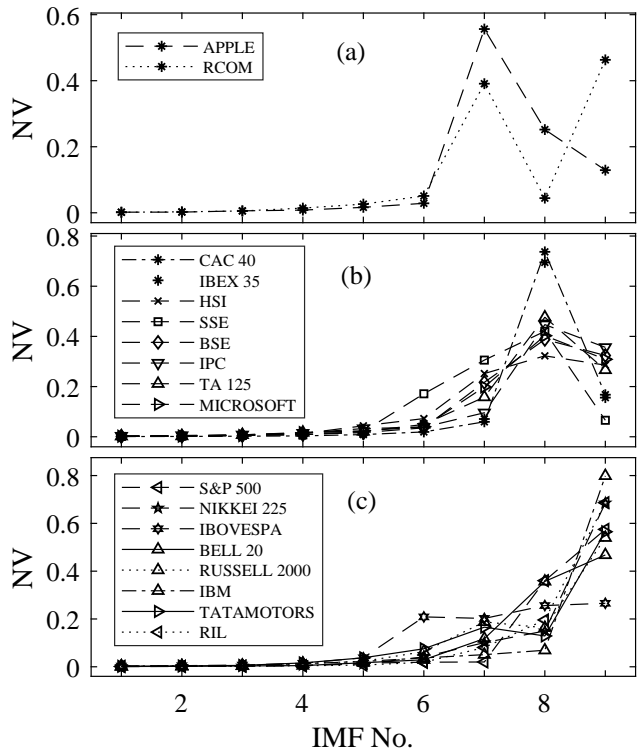}
\caption{\label{fig:Var} Represents the $NV$ of all the IMFs of all the indices and companies. IMF1 to IMF5 $NV$ is very small and $NV$ value increases significantly for IMF6 to IMF9.}
\end{figure}

In Figs.~\ref{fig:Var}(a)-(c) represents $NV$ of all the IMFs of all the indices and companies, where plots have been arranged according to the order of higher $NV$ of IMFs. Figs.~\ref{fig:Var}(a)-(c) show that $NV$ is very low for all the indices and companies up to IMF5, and it increases significantly from IMF6.  Hence $NV$ separates the time series into two time horizons: Short-term time horizon (IMF1 to IMF5) and Long-term time horizon (IMF6 to IMF9) which consistent with the $H$ exponent analysis given above. Fig.~\ref{fig:Var} (a), Fig.~\ref{fig:Var} (b) and Fig.~\ref{fig:Var} (c) show that the $NV$ is higher for IMF7, IMF8 and  IMF9 respectively for the companies mentioned in the figures. The higher value of $NV$ some of IMFs in long term indicate that they may play important role in signal reconstruction~\cite{chatlani2012emd}. 

In summary, we have shown using EMD based Hurst exponent analysis and NV techniques that the market is random in short-term ITH and deterministic in long-term ITH. The time scales for short-term ITH are from few days to 3 months and for long-term ITH is more than 3 months to several years. Two time series $X_{ST}$ and $X_{LT}$ for short-term ITH and long-term ITH have been constructed that can be used to simulate market dynamics in two investment horizons. $X_{ST}$ is random in nature, whereas, $X_{LT}$ is positively correlated with the fundamentals of the company. These results may be very useful for making investment decision in both short-term ITH and long-term ITH. 

\begin{acknowledgments}
We would like to acknowledge Jean-Philippe Bouchaud for some valuable comments and suggestions.
\end{acknowledgments}
\bibliography{apssamp}
\end{document}